\documentclass{svproc}

\usepackage{graphicx,tabularx}
\usepackage{multirow}
\usepackage{amsmath,amssymb,braket,upgreek,bm,verbatim}
\usepackage{xcolor}
\usepackage[absolute]{textpos}
\usepackage{xcolor}
\usepackage[final]{listings}
\usepackage{mdwlist}
\usepackage{framed, color}
\definecolor{mygray}{rgb}{0.5,0.5,0.5}

\lstdefinestyle{my-c-numbered}{
  language=C,
  showspaces=false,
  columns=fullflexible,
  breaklines=true,
  frame=single,
  emphstyle=\bf,
  basicstyle=\footnotesize,
  stringstyle=\ttfamily,
  breaklines=true,
  numbers=left,
  numberstyle=\tiny,
  stepnumber=1
}

\lstdefinestyle{my-bourne-numbered}{
  language=bash,
  showspaces=false,
  columns=fullflexible,
  breaklines=true,
  frame=single,
  emphstyle=\bf,
  basicstyle=\footnotesize\ttfamily\scriptsize,
  commentstyle=\itshape\color{mygray},
  stringstyle=\ttfamily,
  breaklines=true,
  numbers=left,
  numberstyle=\tiny,
  stepnumber=1
}

\lstdefinestyle{my-docker-numbered}{
  language=bash,
  showspaces=false,
  columns=fullflexible,
  breaklines=true,
  frame=single,
  emphstyle=\bf,
  basicstyle=\footnotesize\ttfamily\scriptsize,
  commentstyle=\itshape\color{mygray},
  stringstyle=\ttfamily,
  breaklines=true,
  numbers=left,
  numberstyle=\tiny,
  stepnumber=1
}

\lstdefinestyle{my-xml-numbered}{
    language=XML,
    showspaces=false,
    numbers=left,
    numberstyle=\tiny,
    basicstyle=\footnotesize\ttfamily\scriptsize,
    stringstyle=\ttfamily,
    showstringspaces=false,
    keywordstyle=\scriptsize,
    emphstyle={\color{Cyan}},
    morecomment=[s]{<!--}{-->},
    commentstyle=\itshape\color{mygray},
}

\usepackage{url}

\newcommand{\changebegin}[1]{\marginpar[\hspace*{-60pt}\mbox{\hspace*{10pt} 
 $\top$ \tiny ({#1})}]{\mbox{$\top$ \tiny ({#1})}}}
\newcommand{\changeend}[1]{\marginpar[\hspace*{-60pt}\mbox{\hspace*{10pt} 
 $\bot$ \tiny ({#1})}]{\mbox{$\bot$ \tiny ({#1})}}}
\makecompactlist{compactenumerate}{enumerate}
\makecompactlist{compactitemize}{itemize}
\makecompactlist{compactdescription}{description}

\newcounter{todocounter}

\usepackage{wrapfig}

\newif\iffinal

\iffinal
  \newcommand\nagi[1]{}
  \newcommand\chapman[1]{}
  \newcommand\lukens[1]{}
  \newcommand\raphael[1]{}
  \newcommand\nick[1]{}
  \newcommand\muneer[1]{}
  
\else
  
  \newcommand\nagi[1]{{\color{blue}[Nagi: #1]}}
  \newcommand\chapman[1]{{\color{purple}[Chapman: #1]}}
  \newcommand\lukens[1]{{\color{purple}[Lukens: #1]}}
  \newcommand\raphael[1]{{\color{red}[Raphael: #1]}}
  \newcommand\nick[1]{{\color{red}[Nick: #1]}}
  \newcommand\muneer[1]{{\color{red}[Muneer: #1]}}
  
\fi

\begin{document}
\mainmatter              
\title{Lessons Learned on the Interface between Quantum and Conventional Networking}

\author{Muneer Alshowkan, Nageswara~S. V. Rao, Joseph~C. Chapman, Brian~P.~Williams, Philip~G. Evans, Raphael~C. Pooser, Joseph~M. Lukens, and Nicholas~A. Peters}

\institute{Oak Ridge National Laboratory\footnote{This 
            manuscript has been
            authored by UT-Battelle, LLC, under contract DE-AC05-00OR22725
            with the US Department of Energy (DOE). The US government
            retains and the publisher, by accepting the article for
            publication, acknowledges that the US government retains a
            nonexclusive, paid-up, irrevocable, worldwide license to publish
            or reproduce the published form of this manuscript, or allow
            others to do so, for US government purposes. DOE will provide
            public access to these results of federally sponsored research
            in accordance with the DOE Public Access Plan
            (https://energy.gov/downloads/doe-public-access-plan).} \\
Oak Ridge, TN 37831, USA}
\authorrunning{M.~Alshowkan \emph{et al.}}
\titlerunning{Interface between Quantum and Conventional Networking}

\maketitle              

\begin{abstract}
The future Quantum Internet is expected to be based on a hybrid architecture with core quantum transport capabilities complemented by conventional networking.
Practical and foundational considerations indicate the need for conventional  control and data planes that (i)~utilize extensive existing telecommunications fiber infrastructure,
and (ii)~provide parallel conventional data channels needed for quantum networking protocols. 
We propose a quantum-conventional network (QCN) harness to implement a new architecture to meet these requirements. 
The QCN control plane carries the control and management traffic,
whereas its data plane handles the conventional and quantum data communications.
We established a local area QCN connecting three quantum laboratories over dedicated fiber and conventional network connections.
We describe considerations and tradeoffs for layering QCN functionalities, 
informed by our recent quantum entanglement distribution experiments conducted over this network.
\keywords{quantum network, efficient networks, conventional network, quantum key distribution, control plane, teleportation, entanglement distribution}
\end{abstract}

\section{Introduction}

Quantum networks promise fundamentally new capabilities for scientific discovery empowered by accelerator and reactor facilities, quantum computing, and cybersecurity. The highly anticipated Quantum Internet (QI)~\cite{Kimble2008} is expected to enhance the role of the conventional Internet by quantum connectivity with integrated security. Indeed, quantum networks are known to play an important role in secure communications~\cite{Gisin2007,Bennett2014}, distributed computing~\cite{Cirac1999}, blind computing~\cite{Broadbent2009,Barz2012}, and enhanced sensing~\cite{Bollinger1996,Giovannetti2004,Giovannetti2011}.
Their full potential, however, can only be realized by focused research and development efforts in novel quantum network technologies including quantum repeaters, entanglement sources, photon detectors, and powerful infrastructures with capabilities to test, deploy, and transition to production environments.

Practically and fundamentally, \emph{quantum} networking is inextricably tied to \emph{conventional} networking: (i) conventional fiber infrastructure and control-plane technologies are critical to support quantum network deployments, since the construction of a separate quantum infrastructure will likely prove prohibitively expensive; and (ii) all quantum networking protocols require some form of classical communications for their execution. Thus 
both networking capabilities must operate in concert to realize a successful QI.

In this work, we propose a quantum--conventional network (QCN) harness that enables a quantum network to operate its devices in concert with a conventional network, using a control plane for management and orchestration of services
and a data plane as core for quantum information transfer. 
We describe an implementation of a QCN harness over a state-of-the-art quantum network connecting three quantum laboratories at Oak Ridge National Laboratory (ORNL). We summarize entanglement distribution experiments conducted over its connections provisioned using a wavelength-selective switch (WSS).

Our network spanning three buildings uses eight independent entanglement channels (16 wavelengths), in the lowest loss telecommunications band, that are remotely and dynamically reconfigurable. This use case provides a rich scenario to explore the conventional networking needs of a deployed quantum network both in terms of the overall architecture and individual component capabilities and their implementations.

The organization of this paper is as follows. First, we lay out our framework for the network architecture of QCN harness in Section~\ref{sec:framework}. Then in Section~\ref{sec:usecase}, we present a scientific use case that requires QCN control plane to assist the quantum network. Following that, we discuss our prototype network deployment and entanglement distribution experiments in Section~\ref{sec:depnet} and lessons learned in Section~\ref{sec:disc}. Finally, we conclude in Section~\ref{sec:conclusion}.

\section{Generic Quantum-Conventional Network Harness}
\label{sec:framework}

Software-Defined Networking (SDN)~\cite{Feamster2014} provides a flexible architecture in which the network control and data traffic are separated, and often carried over separate logical or physical planes. 
In the control plane, configuration signals are distributed from a central unit known as a controller to the network devices, including the routers and switches, to manage the data plane that carries the network traffic.
The network devices feature a standardized programming interface for monitoring and forwarding network traffic based on the controller's choices. 
Because the control and data planes are separated, the management process in SDN networks is modular and streamlined, thereby simplifying 
network administration and security policy enforcement.

There is a natural separation in quantum networks between the quantum data being handled and the classical control and management signals required for the quantum services, which lends itself to an architecture similar to SDN.
Furthermore, quantum networks require many specialized components at present, including nonclassical photon sources and single-photon detectors, that do not readily interface with conventional network components. For greater utilization of these resources, a dynamic network configuration can provide the flexibility needed to establish quantum services between several nodes, enabling the data plane to be reconfigured temporarily to provide a particular service. Moreover, this approach can be used for network segmentation to offer different network services to different network nodes using the same resources. A similar approach is currently being constructed in quantum computing services, where dedicated access to quantum computers is offered on the cloud.

In general, an SDN network consists of three planes: data, control, and application. 
The \textbf{data plane} is the lowest plane and comprises network equipment, including routers and switches (physical and virtual) and access points. 
Data plane devices are most commonly maintained and accessed via the SDN Control-Data-Plane Interface (CDPI) available in OpenFlow~\cite{McKeown2008} via classically encrypted connections, such as Transport Layer Security (TLS).

The \textbf{control plane} is the middle plane between the data and application planes, consisting of a group of software-based SDN controllers that enable network forwarding operations via the CDPI. 
It enables communications between different controllers and the Application-Control-Plane Interfaces (ACPIs).
There are two main elements in a controller, which are either  functional or logical. The functional component includes coordination and virtualization, and the logical component converts the application's requirements to executable instructions.
The \textbf{application plane} is the top plane in the SDN architecture, comprising one or more applications that interact with the controllers to make internal decisions via an abstract representation of the network.

Our proposed QCN architecture builds extensively on the principles of SDN, whose well-defined planes not only provide a direct path to establish quantum services within the existing Internet Protocol (IP) infrastructure, but also enable flexible allocation and connection of heterogeneous quantum systems.  
In particular, the ability to configure the data plane for routing allows us to segment the QCN design and create customized services in the same network. 
QCNs are expected to be composed of (i) quantum devices such as switches, repeaters, photonic qubit sources, quantum processing or computing systems, and other end-node measurement systems; (ii) quantum links that support communications within and between these devices over local- and wide-area connections; (iii) conventional networking for classical data and control transport; and possibly (iv) classical analog control signals for quantum-device management.

From an infrastructure perspective, the QCN data plane is where quantum communications take place.
Indeed, the quantum part of the QCN data plane is somewhat akin to the dedicated optical data plane in conventional IP networks~\cite{Runseretal2007}.
To be effective in U.S. Department of Energy (DOE) science infrastructures, such a quantum data plane must span various geographically distributed sites that house scientific instruments, and also integrate with existing IP networks, including DOE's Energy Science Network (ESnet).
Certainly, dedicated data plane connections have been found to be very effective in provisioning high bandwidth optical connections in
science environments, for example, OSCARS circuits over ESnet \cite{oscars} and dedicated wavelengths over UltraScienceNet (USN) \cite{Raoetal2005commag}. The control plane design in this paper is based on USN, which implemented a cross-country encrypted control plane using firewalls over a separate IP network, about a decade prior to a full SDN framework development in \cite{Feamster2014}.  
\begin{figure}[tb!]
\centering
       \includegraphics[scale=0.35]{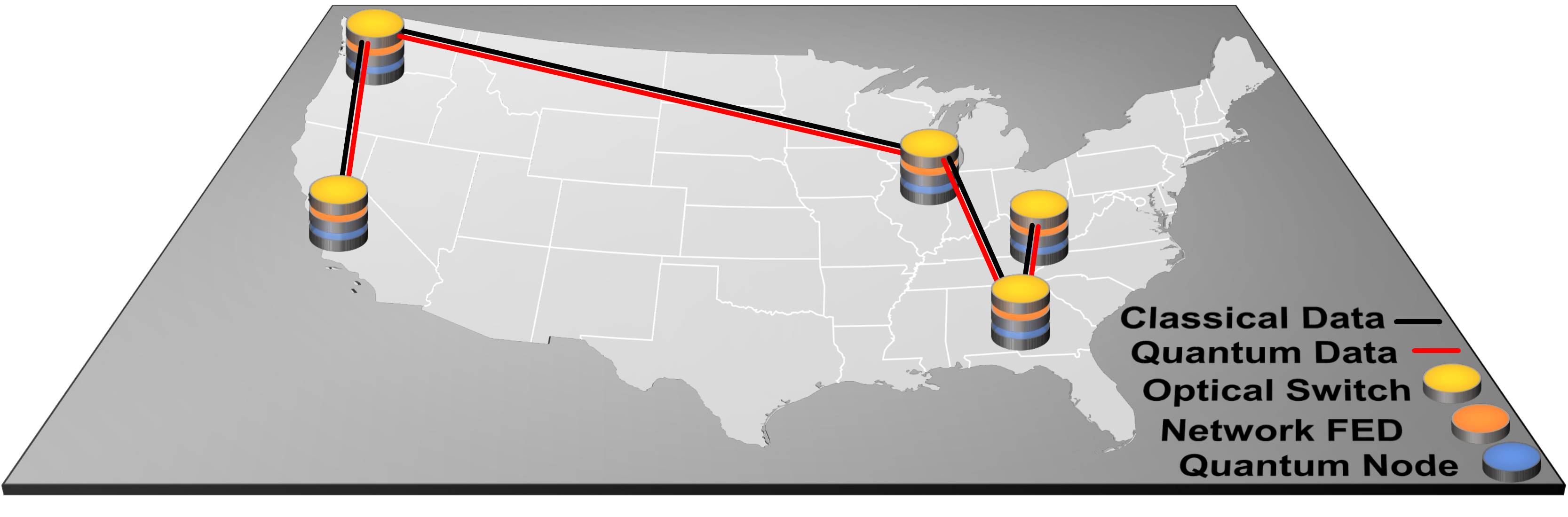} 
  \caption{Conceptual view of a QCN data plane supporting provisioning of conventional and quantum channels. Node locations are illustrative and modeled after USN.}
\label{fig:data-plane}
\vspace*{-0.2in}
\end{figure}

QCN solutions for quantum connections for DOE science environments entail solving challenges including
the development of control (both conventional digital and analog physical layer)  and management planes for quantum devices, as well as scalable frameworks for jointly developing and testing
QCN hardware devices and associated software.
We address both these areas here by building upon previous DOE projects in high-performance networking for science infrastructures that consist of a small number of sites with stringent bandwidth requirements. 
In the next few paragraphs, we provide an overview of the data plane, control plane, and application plane designs envisioned for a QCN. 

{\bf QCN Data Plane:}
The QCN data plane is expected to be composed of quantum and conventional network devices. Conventional network devices include routers, switches, and firewalls---either physical or virtual as in Palo Alto firewall and encryption devices (FED). On the other hand, the quantum part includes all-optical switches, nodes with quantum resources, and (in the future) quantum repeaters, as illustrated in Fig.~\ref{fig:data-plane}. Its architecture and layered stack are specific to quantum signals and protocols as described in Sec. \ref{subsec:prototype}. In particular, quantum computing systems may be connected over quantum links to support quantum computations over a distributed quantum computer, and furthermore, they can be directly connected to a storage system comprised of quantum and classical memories.
For quantum data transfers, these connections promise inherent security since eavesdropping on quantum channels is precluded by the no-cloning theorem~\cite{Wootters1982}. Further, quantum key distribution (QKD) can be used with one-time-pad encryption to protect \emph{classical} data transfers as well~\cite{Runseretal2007}. By integrating computing, communications, and storage devices, a distributed quantum computer, with each function carried out by specialized hardware, may be able to attain performance well beyond that possible with a single quantum computer---not unlike current high-performance computing architectures based on interconnected GPUs and CPUs. Furthermore, the required quantum connections between these devices can be established effectively using control plane technologies that exploit the softwarization of quantum devices~\cite{RaoHumble2018ascr}, further described below. In the ultimate case of a continental quantum network, the QCN data plane will consist of repeaters in the core which are connected over fiber paths to other repeaters and end nodes.

\begin{figure}[tb!]
\centering
    \includegraphics[scale=0.35]{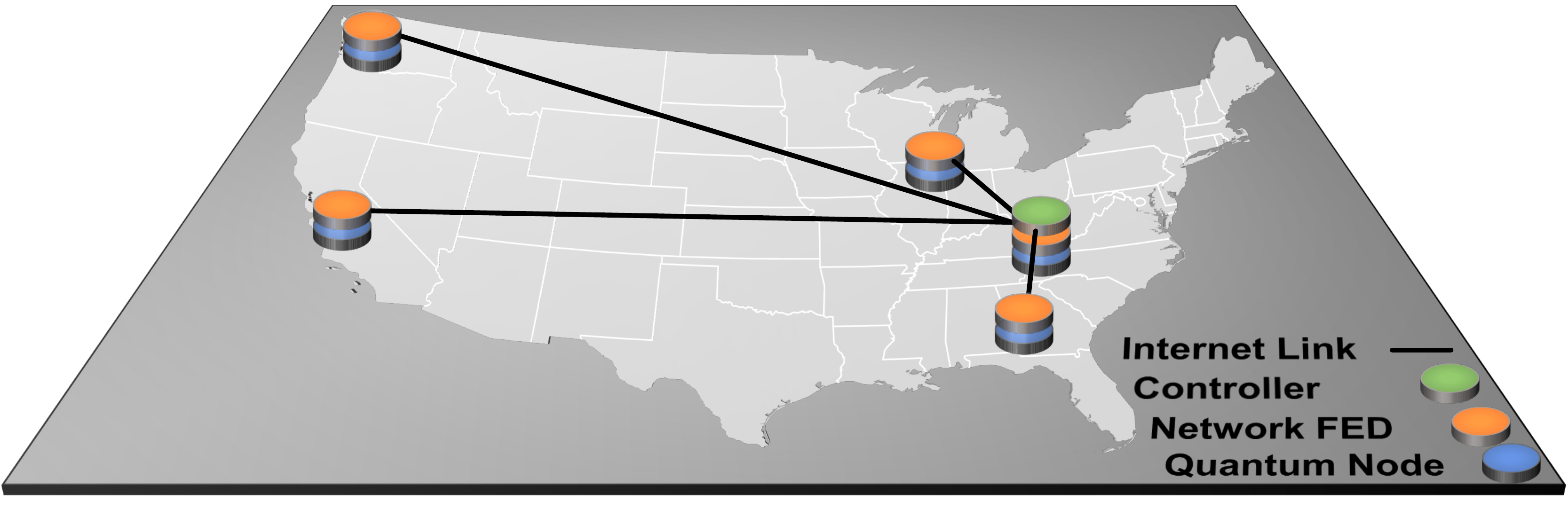} 
  \caption{The QCN control and management plane provides wide-area connections for configuration and management. (Node locations modeled after USN.)}
\label{fig:control-plane}
\end{figure}

{\bf QCN Control Plane:}
As noted above, recent advances in SDN control and management plane technologies~\cite{Aguado2020,Humbleetal2018} can be leveraged to provision quantum connections between nodes on demand. 
In USN specifically, a separate control plane was deployed to configure high-bandwidth optical connections between sites, wherein encrypted tunnels between firewalls carried the control and management traffic. Similar control planes can be developed for quantum networks to protect against cyberattacks while providing the security features of authentication and encryption using FEDs \cite{Raoetal2005commag} as shown in Fig.~\ref{fig:data-plane}, ideally secured with quantum keys. 
This solution requires softwarization wrappers for quantum devices so that they can be managed using SDN technologies over the IP control plane. One such method is based on using Virtual Machines (VMs) to implement an Open vSwitch as a frontend and custom communication channels to a quantum switch such that it can be controlled through OpenFlow. Such a solution for local connections is described in \cite{Humbleetal2018}. This approach is applicable to generic quantum connections, and additionally provides security features needed for wide-area deployments using  firewall and encryption devices.  

{\bf QCN Application Plane:}
Even though there is not yet widespread quantum network deployment, there are already several proposed quantum applications which rely on network-aware capabilities. For example, quantum-enhanced telescopy~\cite{Lukin1,Lukin2} will need entanglement provisioning for its memories which will be dependent on the image brightness and will have precision timing delay requirements that will vary from other applications. Thus, we propose a quantum application plane on top of the secure control plane for the purpose of providing requests and constraints on the control plane for such applications, similar to conventional SDN where the application layer can request network functionality~\cite{sdnarchitectureoverview_2013}. In our prototype, we execute a simple bandwidth provisioning application, testing various bandwidth allocation configurations for our network.

In the following sections, we show through example how the QCN structure introduced here can be applied to the use case of entanglement distribution in a deployed QLAN, highlighting the challenges encountered and lessons learned in our initial research.

\section{Scientific Use Case}
\label{sec:usecase}

\subsection{Entanglement Distribution}

One of the most basic quantum networking capabilities is the support of on-demand entanglement between multiple parties.
Currently, quantum network implementations can be classified at the logical level into four categories: point-to-point~\cite{Scarani2009}, trusted-node~\cite{Elliott2002,Peev2009,Stucki2011,Sasaki2011,Chen2010,Wang2014,Mao2018,Dynes2019,Evans2019}, point-to-multipoint~\cite{Townsend1997,Lim2008}, and fully connected~\cite{Wengerowsky2018,Joshi2020,Lingaraju2020a,Appas2021}. 

For quantum networks to be practical and useful, they must have nodes that are spatially separated and independent, ideally supporting heterogeneous quantum resources (stationary qubits, detectors, photon sources, etc.).
Furthermore, the network architecture needs to be suitable for interconnecting with other network topologies to form larger, more complex networks. Classical networking capabilities, e.g., a control plane for management and a parallel data plane for node-to-node classical communications, will also be needed.  
Fully connected entanglement networks with dense wavelength-division-multiplexed allocation show promise but have been limited so far to demonstrations where all detection events occur at the same physical site~\cite{Wengerowsky2018,Joshi2020,Lingaraju2020a,Appas2021}. While an important preliminary step, the time synchronization and data management requirements are, in these cases, excessively simplified: high-quality local connections or fiber loopbacks sidestep difficult issues that arise when 
detection systems are tied to different local clocks, for example.
Our prototype testbed is designed to address the above needs by leveraging the QCN architecture to enable a dynamic, fully connected QLAN 
with adaptive bandwidth provisioning and simultaneous remote detection.

\label{sec:prototype}

\subsection{Prototype Network Architecture}
\label{subsec:prototype}

\begin{figure*}[tb!]
	\centering
	\includegraphics[width=\textwidth]{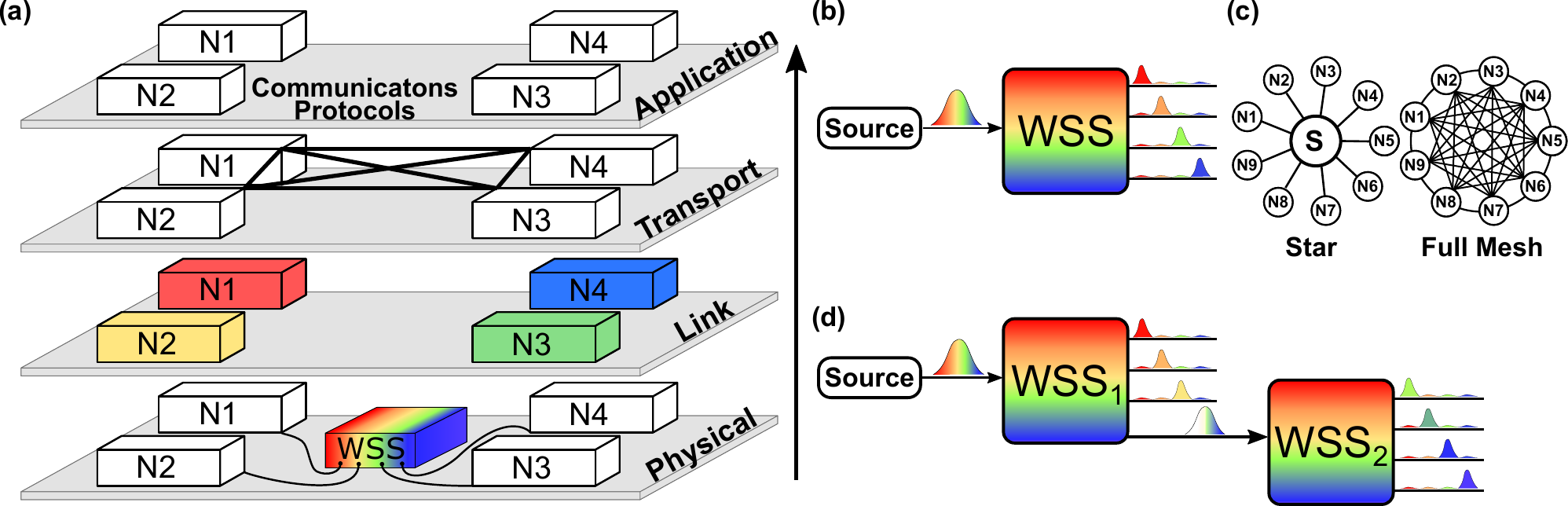}
	\caption{(a) QCN quantum data plane layers and services. The physical layer includes the optical components where the quantum photons travel and are manipulated, the link layer slices the spectrum and routes each slice to a particular user, the transport layer is the quantum-correlated network where a pair of users shares entanglement, and the application layer uses the entangled pairs to perform a service. (b) Single WSS configuration: the input spectrum is sliced and routed to the output ports. (c) Comparison of physical (star) and logical (fully connected mesh) topologies in network design. (d) Nested WSS where a portion of the spectrum is routed from the first to the second WSS, thereby expanding the number of nodes connected by the entanglement source~\cite{alshowkan2021}.}
	\label{fig2}
\end{figure*}

Our vision for a layered network stack, introduced in \cite{alshowkan2021}, is outlined in Fig.~\ref{fig2}(a). Drawn from the Transmission Control Protocol/Internet Protocol (TCP/IP) stack~\cite{Cerf1974,braden1989}, this network stack includes not only the physical medium and routing but also the protocols and applications which define the node behavior in the network. Our stack specifically leverages spectral multiplexing, so that the physical layer consists of fiber links connected with WSSs. According to the QCN control and QCN data plane distinction described in Sec.~\ref{sec:framework},  this layered stack can be viewed primarily as an elaboration of the quantum part of the QCN data plane that carries quantum signals using quantum protocols; by contrast, the QCN control plane and the conventional part of the QCN data plane follow the conventional TCP/IP stack.

Thus in our case, for example, the instructions sent to the WSS for configuration fall under the QCN control plane, while the photon detection timestamps are the purview of the conventional networking part of the QCN data plane. 

In this architecture, each node/user is directly connected to the WSS by deployed fiber [Fig.~\ref{fig2}(b)], corresponding to the star topology in Fig.~\ref{fig2}(c). The link layer uses the WSS to partition the bandwidth of the entangled photon source and dynamically route the allocated spectral channels to their assigned user. The transport layer is characterized by a logical fully connected mesh topology shown in Fig.~\ref{fig2}(c). In this way, the nonlocal potential of quantum entanglement provides greater flexibility for network topologies beyond restrictions of the physical layer alone. Indeed, as shown here, a star \emph{physical} topology (all $N$ nodes connected to a central source) can result in a fully connected \emph{logical} mesh that gives entanglement connections between all $N(N-1)/2$ user pairs. These logical connections are then used in the application layer for quantum services/protocols between the nodes/users, e.g., QKD, quantum teleportation, and remote state preparation.

In our stack, we have purposely omitted an internet layer for connecting several networks since this is not needed  in our single-network testbed to illustrate the QLAN concept. However, building on previous work for quantum access networks~\cite{Ciurana2015,Alshowkan2016}, this architecture could be expanded to include multiple nested WSSs [Fig.~\ref{fig2}(d)] 
for future connectivity between multiple QLANs in larger, more complex networks. Regardless, quantum networks are in their early stages, so any network stack we propose is indeed tentative, with future refinements and modifications expected as development continues. 

\section{Deployed Network}
\label{sec:depnet}
\subsection{Time Synchronization}
\label{subsec:time}

Distributed timing synchronization is crucial for extending tabletop quantum networking experiments to the field. 
Timing requirements for quantum networks are usually much stricter than for conventional LANs. In particular, the ubiquitous standard for synchronization over ethernet, the precision time protocol (PTP), is designed to attain only sub-$\upmu$s jitter~\cite{IEEE1588}, whereas sub-ns precision is often sought for distributed photon-counting experiments. White Rabbit~\cite{Linpinski2011}, a major improvement to PTP that uses synchronous ethernet to reach ps-level precision, thus offers considerable promise for future quantum network designs. 
Alternatively, previous quantum communications experiments have used fully optical approaches to compensate for clock drift between nodes, including direct tracking of pulsed quantum signals ~\cite{Hughes2005,Chapuran2009} or photon coincidence peaks~\cite{Peloso2009,Krenn2015,Steinlechner2017,Shi2020}.

\begin{figure}[tb!]
	\centering
	\includegraphics[scale=0.4]{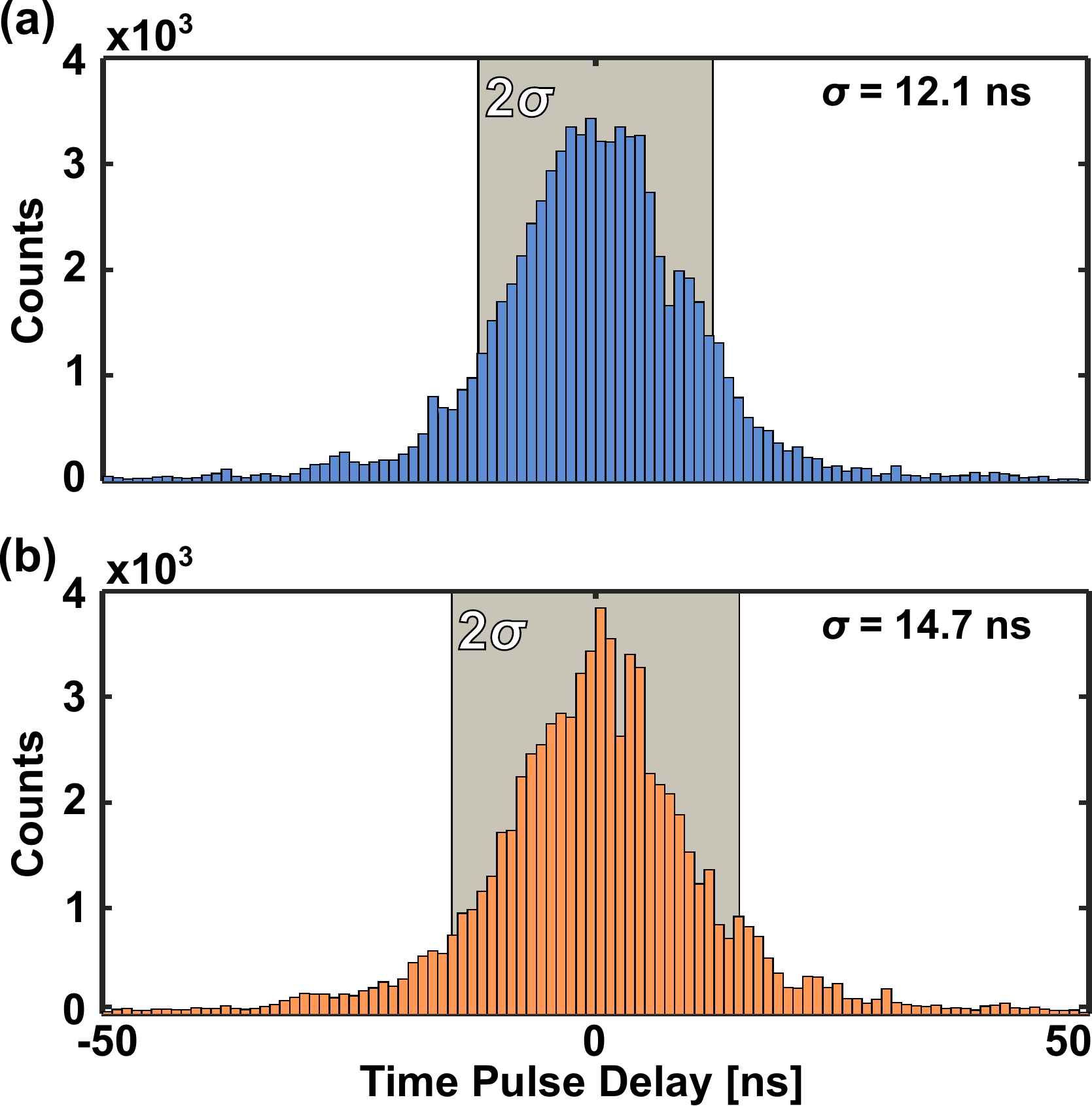}
	\caption{Histogram of relative delays between time pulses from two GPS receivers~\cite{alshowkan2021}. (a) Alice and Bob. (b) Charlie and Alice.}
	\label{jitter}
\end{figure}

For our QLAN~\cite{alshowkan2021}, we use synchronization based on the Global Positioning System (GPS) due to its cost-effectiveness, simplicity, and availability (a GPS antenna was already located in one lab). The GPS signals allow us to derive two clocks from our Trimble Thunderbolt E receivers: a 10~MHz tone and a pulse per second (PPS). Multi-hour characterization of the timing jitter for independent receivers  
returned the histograms shown in Fig.~\ref{jitter}. The relative-delay distributions have standard deviations of about 10~ns, which is a significant improvement over PTP.

\label{What need to be solved}

\begin{figure*}[tbh!]
	\centering
	\includegraphics[width=\textwidth]{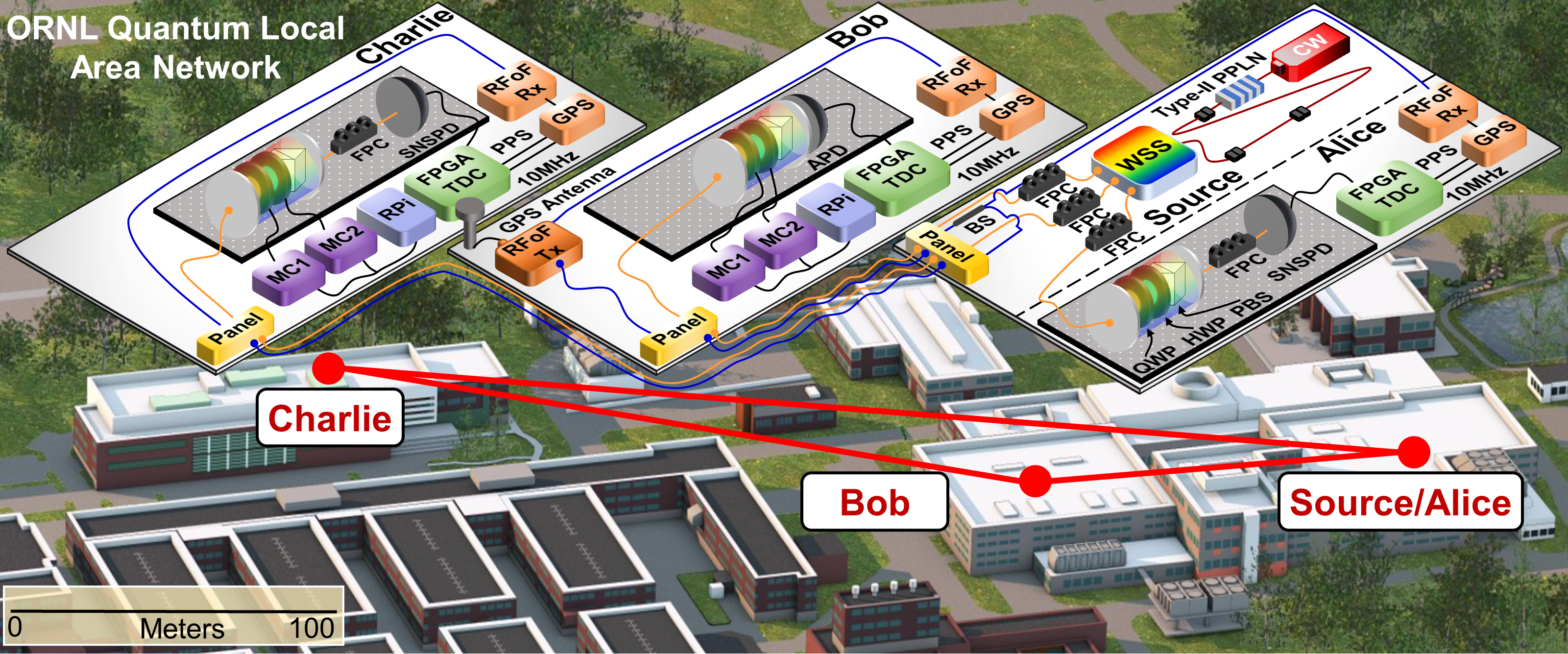}
	\caption{Map of QLAN on ORNL campus~\cite{alshowkan2021}. The receiver configurations at each node are shown as insets. APD: avalanche photodiode. CW: continuous-wave laser. FPC: fiber polarization controller. FPGA: field-programmable gate array. GPS: Global Positioning System. HWP: half-wave plate. MC: motion controller. Panel: fiber-optic patch panel. BS: beamsplitter. PBS: polarizing beamsplitter. PPLN: periodically poled lithium niobate. PPS: pulse per second. Source: entangled photon source. QWP: quarter-wave plate.  RFoF Rx: RF-over-fiber receiver. RFoF Tx: RF-over-fiber transmitter. RPi: Raspberry Pi microprocessor board (to control MCs). SNSPD: superconducting-nanowire single-photon detector. WSS: wavelength-selective switch.}
	\label{setup_map}
\end{figure*}

\subsection{Experimental Implementation}

Our prototype network is set up across three buildings at ORNL, as shown in Fig.~\ref{setup_map}. 
The entangled photon source used is a fiber-coupled periodically poled lithium niobate (PPLN) crystal phase-matched for type-II spontaneous parametric down-conversion (SPDC)~\cite{alshowkan2021}. The state produced is then split into frequency bands which are distributed to the network users. Using the WSS, we partition 8 pairs of frequency-correlated channels, each having bandwidth $\Delta\omega/2\pi=25$~GHz and aligned to the International Telecommunication Union (ITU) grid (ITU-T Rec. G.694.1). The channels are centered at $\omega_n = \omega_0 \pm \Delta\omega(n-\frac{1}{2})$ for the signal (idler). These channels span $\sim$3~nm in the C-band (1557.3--1560.5~nm).

Alice and the entangled photon source are within the same lab; meanwhile Bob and Charlie are in separate buildings, each connected to the source through approximately 250~m and 1200~m fiber path lengths, respectively. The WSS connects the entangled photon source and each user by a direct fiber link. The WSS outputs directed to fiber-based polarization controllers (FPC) for passive polarization drift compensation: Alice's FPC output goes directly to her polarization analyzer, while Bob's and Charlie's connections include inter-building transmission fibers. The polarization analyzers are connected to single-photon detectors; Alice and Charlie are connected to highly efficient (but polarization-dependent)  single photon detectors and Bob is connected to an InGaAs avalanche photodiode (APD).

Although each user could use their own GPS antenna, it is simpler and cheaper for us to distribute the GPS signal from Bob's location to the other nodes using RF over fiber (RFoF). Using the 10~MHz and 1~Hz (PPS) clocks, we synchronize the FPGA-based time-to-digital converter (TDC) at each node, which bins the photon detection events at 5~ns resolution according to the FPGA's internal 200~MHz clock. This approach is scalable for increasing network size since the resources can be readily duplicated for additional nodes.

 \begin{figure}[tb!]
 	\centering
	\includegraphics[scale=0.45]{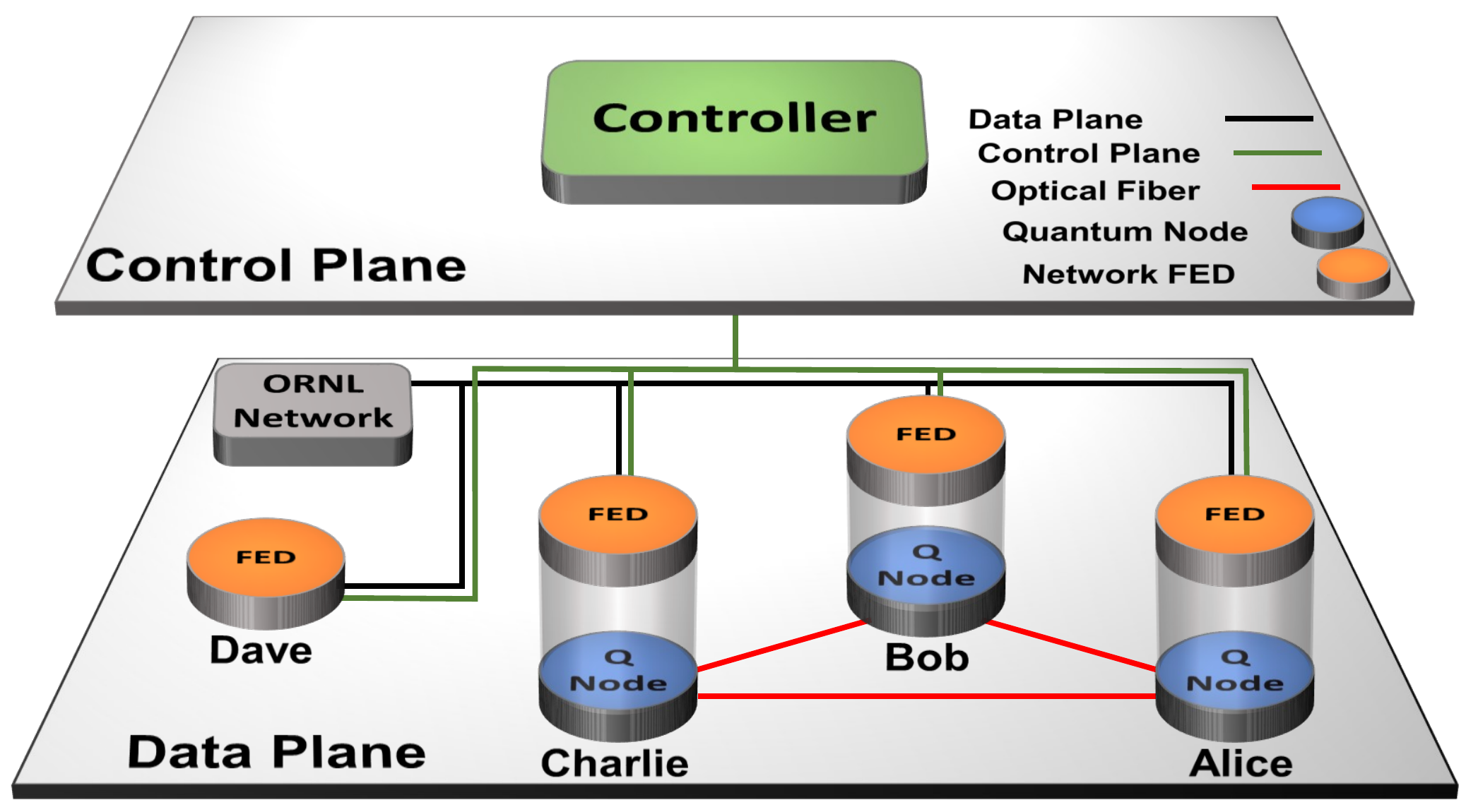}
	\caption{Overview of the QCN control and data planes for the ORNL QLAN. Commands from the QCN control plane define routing connections for both the ethernet (black lines) and optical (red lines) traffic of the QCN data plane, the former carrying digital information such as timestamps while the latter carries photonic qubits.}
	\label{fig:qlan-control-plane}
\end{figure}

\begin{figure}[tb!]
 	\centering
 	\includegraphics[trim=0cm 0.9cm 0cm 0cm,width=0.6\textwidth]{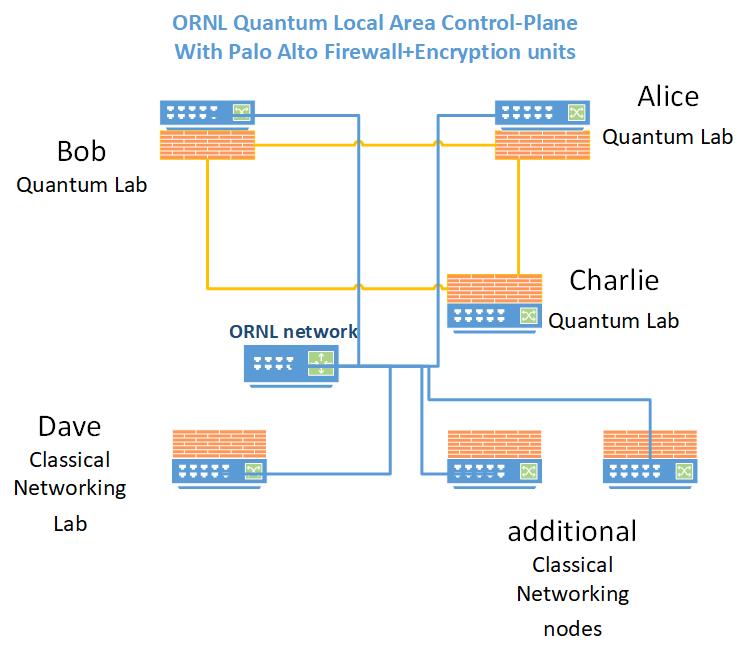}
	\caption{Implementation of QCN conventional control and data palens using Palo Alto 220 FEDs. Secure tunnels (orange lines) between three quantum labs provide conventional network connectivity to support quantum experiments. Multiple conventional subnets are supported on secure sides of FEDs, which are in turn connected over the secure tunnels.
	The virtual router and timing exchange devices of quantum nodes are connected to these subnets for control and data conventional communications.}
	\label{fig:qlan-control-plane-implementation}
\end{figure}

To manage the measurement and instrument information flow between nodes, we employ a QCN control plane that establishes 
 
encrypted tunnels between the three quantum labs (Alice, Bob, and Charlie).  
As summarized in Fig.~\ref{fig:qlan-control-plane}, a virtual router in each device enables each node to communicate with the other devices over these QCN data and control planes. 
QCN conventional control and data planes are implemented using
Palo Alto 220 FEDs as shown in Fig.~\ref{fig:qlan-control-plane-implementation}, that provide 
encrypted tunnels between the three quantum labs (Alice, Bob, and Charlie) and the conventional networking node (Dave).
Multiple conventional subnets are supported behind each FED, to which devices including instruments, microprocessors, computing systems, and TDCs are connected.
In the QCN data plane, photon detection events are time-tagged by the TDCs and then transferred to a central computer at Alice for coincidence counting and analysis.
  
To ensure experimental integrity, it is critical that the conventional communications of the QCN data plane have access to sufficient 
bandwidth to handle the conventional data needs of the quantum network. For example, our photon counting experiments produce up to about 3~Mb/s of 32-bit timestamps which are handled easily by our $\sim$1~Gb/s conventional data plane. 

\subsection{Bandwidth Allocation}
Our eight-channel polarization-entangled photon source allows for a plethora of bandwidth allocations that can be optimized for a desired network configuration. The WSS enables dynamic real-time bandwidth provisioning without changing any fiber connections. Entanglement distribution with this approach was realized recently in a tabletop experiment~\cite{Lingaraju2020a}, and here we summarize some of the key results extending this approach to our deployed QLAN~\cite{alshowkan2021}.

The link efficiencies are highly imbalanced due to different deployed-fiber link loss and heterogeneous detector technology. In decreasing order of  combined efficiency are Charlie and Alice (C--A), Alice and Bob (A--B), and Bob and Charlie (B--C). For Allocation 1, we choose to balance entanglement distribution rates as closely as possible, which in our case amounts to assigning the lowest-flux Ch. 8 to C--A, the highest flux Ch. 1 to A--B, and the remaining Ch. 2--7 to B--C. For Allocation 2, we seek to improve the average state fidelity between all channels, which can be degraded due to multipair emission and frequency-dependent polarization transformations. Specifically, we assign Ch. 1--2 to B--C, Ch.~3 to A--B, and Ch. 4 to C--A, which leaves the remaining channels available for future network expansion.

\begin{figure*}[tb!]
	\centering
	\includegraphics[width=0.85\textwidth]{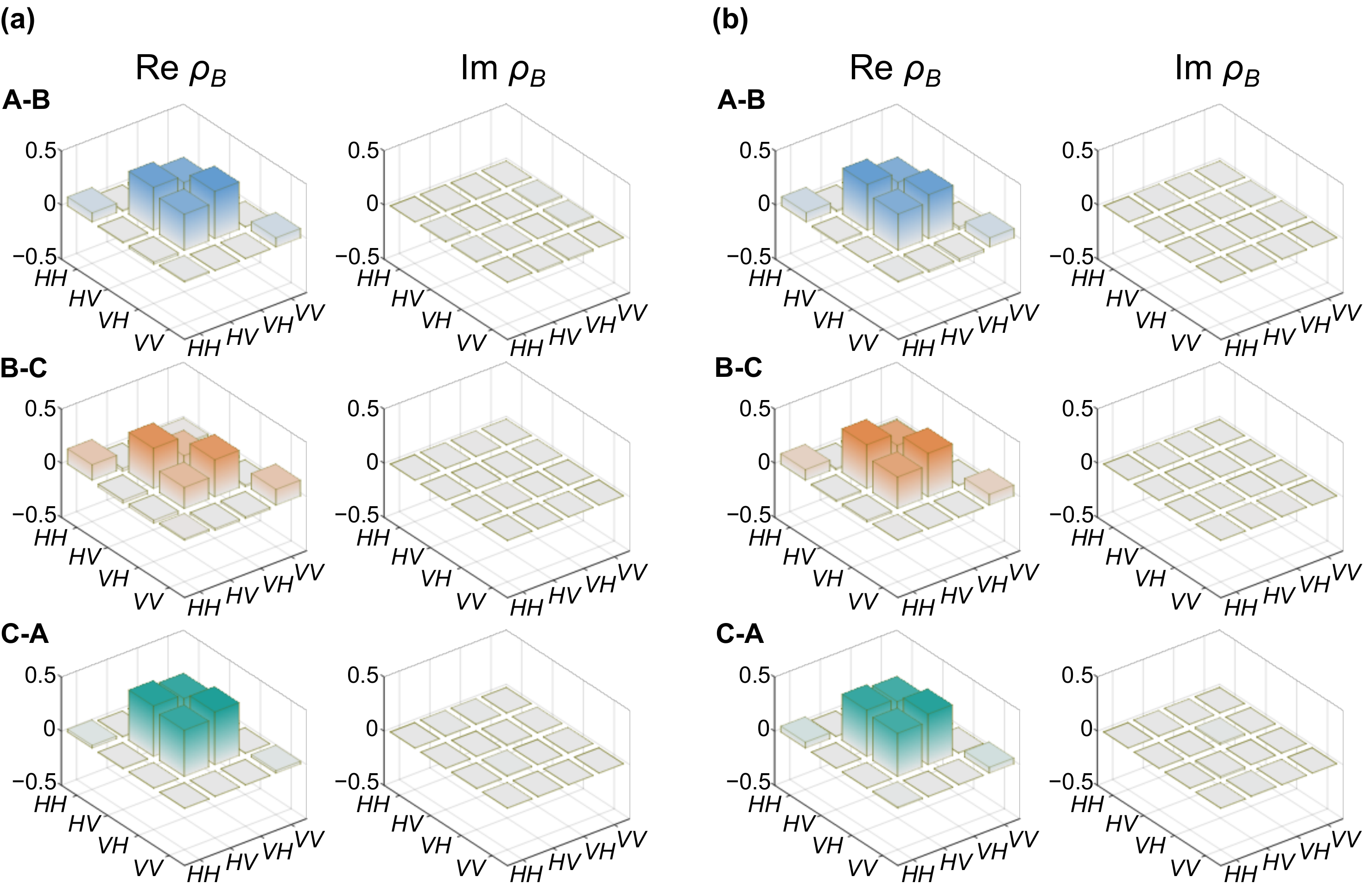}
	\caption{Density matrices estimated by polarization tomography for each pair of users~\cite{alshowkan2021} for (a) Allocation 1 and (b) Allocation 2.}
	\label{alloc_density}
\end{figure*}

Figure~\ref{alloc_density} shows results of polarization state tomography for all pairs of users under these allocations, using a 10~ns coincidence window, 60~s integration time, and Bayesian mean estimation~\cite{Lukens2020b}. These measurements were conducted with only the logical link being measured passing through the WSS which provides equivalent results to spectrally resolved detection. The state fidelities with respect to the ideal maximally entangled Bell state  are shown in Table~\ref{alloc}; these data are without accidental coincidence subtraction.

\begin{table}[b!]
	\centering
	\caption{Link data for both bandwidth allocations~\cite{alshowkan2021}.}
	\label{alloc}
	\begin{tabular}{c|c|c|ccc}
		\textbf{Alloc.} & \textbf{Link} & \textbf{Ch.} & \textbf{Fidelity} & $\bm{E_\mathcal{N}}$ \textbf{[ebits]} & $\bm{R_E}$ \textbf{[ebits/s]} 		\\ \hline
		\multirow{3}{*}{1} 
		& A--B  &  1 	&  $0.75 \pm 0.03$ & $0.70 \pm 0.08$ & $56  \pm 6$ 	\\
		& B--C  &  2--7 &  $0.55 \pm 0.06$ & $0.4 \pm 0.2$ & $30  \pm 10$ \\ 
		& C--A  &  8    &  $0.90 \pm 0.01$ & $0.89 \pm 0.03$ & $206 \pm 6$ 	\\ 	
		\hline
		\multirow{3}{*}{2} 
		& A--B  & 3     &  $0.75 \pm 0.03$ & $0.70 \pm 0.09$ & $57  \pm 7$ 	\\  
		& B--C  &  1--2 &  $0.69 \pm 0.04$ & $0.6 \pm 0.1$ & $26  \pm 4$ 	\\ 
		& C--A  &  4    &  $0.84 \pm 0.02$ & $0.82 \pm 0.05$ & $320 \pm 20$ \\
	\end{tabular}
\end{table}

The reconstructed density matrices provide complete information for protocol performance prediction at a per-photon-pair level, but do not offer insight into important aspects such as the rate. 
Therefore, we consider an ``entanglement bandwidth'' metric~\cite{alshowkan2021}, which combines log-negativity $E_\mathcal{N}$~\cite{Vidal2002} and the coincidence rate to provide a measure of the entangled bits (ebits) per second ($R_E$). This metric, along with the supporting log-negativity, is shown for both bandwidth allocations in Table~\ref{alloc}. Fidelity and $R_E$ are not necessarily positively correlated as evidenced by the B--C and C--A links in Allocations 1 and 2 due to the photon flux that $R_E$ also considers. Consequently, optimal bandwidth allocation is dependent on the desired objectives, which are not necessarily known at initial network deployment and are subject to change. Bandwidth provisioning using a WSS provides remarkable flexibility to accommodate future objectives without disrupting network service with physical interconnection changes.


\section{Summary of Lessons Learned}
\label{sec:disc}

Through the process of designing and implementing the QCN paradigm within a deployed QLAN, we learned several important---and in many cases, unanticipated---practical lessons regarding the efficient operation of distributed quantum networks. Here we summarize our findings, classified into three main categories: (i)~wired vs. wireless, (ii)~timing synchronization, and (iii)~security.

\textbf{Wired vs. Wireless:}~The FPGA-based TDCs used in this experiment host a minimal operating system (OS) for limited hardware resources and optimization. Often, it is challenging to integrate them with third-party devices, especially if there is no support from the OS kernel or the manufacturer. Admittedly, this leads to unoptimized devices and limited capabilities, and can result in overall device performance that is highly dependent on the available network.

In general, there exist two options for network interfaces to share data between devices: 

wired via ethernet and wireless over an access point. The wired connection requires a network infrastructure that may not be easy to deploy. Even if a network exists, device OS, model, firmware, and vendors must be approved in the networking policies, which presents unique challenges for research prototypes in particular. The alternative to the wired connection is to set up a wireless access point with greater flexibility and a lower deployment cost. Yet although wireless networking is an attractive solution, 
we found in practice that its implementation led to unreliable data transfer. The lack of specialized wireless hardware support in the  FPGA hardware resulted in a lower transfer rate and loss of timestamps experimentally.

Accordingly, we found that the native and widely supported ethernet connection in our TDCs offered greater reliability for data transfer. Importantly, the QCN control plane enables a transparent connection between the network nodes using existing research network infrastructure---without impacting the business network.
This advantage of the control plane design was one we did not fully appreciate when we began this project: it allows us to realize the speed of wired ethernet while simultaneously shielding our devices from the management requirements of an enterprise network.

\textbf{Timing Synchronization:}~The photon counts collected to validate entanglement between the network nodes require precise timing, which can be subdivided into two categories: nonlocal clock distribution and instruction coordination. The former defines the absolute limits of site-to-site jitter, and was discussed in detail in Sec.~\ref{subsec:time}, whereas the latter proves critical for collecting and organizing data---e.g., matching results with corresponding measurement device settings at each location. While much less stringent from the perspective of raw time, we found this latter point to pose interesting challenges distinct from clock distribution.

Instruction timing to start and stop a measurement play a crucial role in the expected quality of the distributed entanglement. 
Ideal timing helps optimize coincidence counting and reduce the processing time spent comparing out-of-detection window events. In our experiment, we observed issues when network delays prevented a timely arrival of measurement instructions.
Because we depend on the GPS PPS to define a specific measurement period, any network delays greater than a second resulted in a discrete shift of timestamps between two locations, which 
needed to be corrected through additional post-processing. 
The dedicated links between the nodes provided by the control plane enabled us to send parallel instructions to each pair of nodes using script files which 

configured the instruments and helped minimize such network-induced offsets in the time records to 1~s or less.

\textbf{Security:}~Current security measures in conventional networking devices such as firewalls employ public-key encryption. Because exchanging secret keys over long distances is difficult, a digital certificate or pre-shared secret key is used for initial authentication and encryption. Then, session keys are exchanged using public-key cryptography. While Our current QLAN utilizes these principles for securing the control plane firewalls, 
a more secure system can be implemented by utilizing secret keys pulled from QKD. It is possible to configure the control plane firewall to utilize external symmetric keys for encryption and authentication within the same network infrastructure. Upgrading our QLAN to full-QKD security thus forms a major goal of future work to ensure the long-term integrity of all classical information on the QCN.


\section{Conclusion}
\label{sec:conclusion}
We have proposed a method to enable a quantum network to operate its devices in concert with a conventional network, showing how this QCN harness was able to support reconfigurable entanglement distribution in our prototype network.

The QCN control plane harmonizes the network by configuring devices on-demand including end-to-end data transfer, thus maximizing utilization of available quantum resources. Multiple controllers in future designs will be essential to manage different and specialized services analogous commercial quantum computing services on the cloud.
Future work involves assessing and integrating additional synchronization and timing requirements of control-plane QI devices and deployments.

We note that the polarization entanglement used in our prototype is well suited for quantum networks, especially ones with short and environmentally protected fibers that minimally disturb the quantum state over time. Even submarine fibers show enough stability for polarization encoding to be used successfully over $\sim$100~km without active compensation~\cite{Wengerowsky2019}. Finally, it should be possible to extend our work to a larger, more complex network of networks, by subdividing the biphoton spectrum and sending different parts to several WSSs~\cite{alshowkan2021}.

\section*{Acknowledgments}

This research used resources of the Oak Ridge Leadership Computing Facility at the Oak Ridge National Laboratory, which is supported by the Office of Science of the U.S. Department of Energy under Contract No. DE-AC05-00OR22725.  
This research is supported by the U.S. Department of Energy, Office of Science, Office of Advanced Scientific Computing Research, through the Early Career Research Program and Transparent Optical Quantum Networks for Distributed Science Program (Field Work Proposals ERKJ353 and ERKJ355). 

\bibliographystyle{osajnl}

\end{document}